\documentclass[prc,preprint,showpacs,preprintnumbers,amsmath,amssymb]{revtex4}

\usepackage[mathscr]{eucal}
\usepackage{graphicx}
\def\slr#1{\setbox0=\hbox{$#1$}           
   \dimen0=\wd0                                 
   \setbox1=\hbox{/} \dimen1=\wd1               
   \ifdim\dimen0>\dimen1                        
      \rlap{\hbox to \dimen0{\hfil/\hfil}}      
      #1                                        
   \else                                        
      \rlap{\hbox to \dimen1{\hfil$#1$\hfil}}   
      /                                         
   \fi}

\def\kp{k^{\,\prime}}

\def\ksq{k^2}

\def\myint#1{\!\int\!\!\frac{d^4\!{#1}}{(2\pi)^4}\,}
\def\mytint#1{\!\int\!\!\frac{d^3\!{#1}}{(2\pi)^3}\,}
\def\gev#1{ GeV${}^{#1}$}
\def\be{\begin{eqnarray}}
\def\ee{\end{eqnarray}}

\renewcommand{\theequation}%
    {\arabic{section}.\arabic{equation}}
\makeatletter \@addtoreset{equation}{section} \makeatother

\begin{document}


\title{Quark and Nucleon Self-Energy in Dense Matter}

\author{L.S.Celenza}
\author{Hu Li}
\author{C.M. Shakin}
 \email[email:]{casbc@cunyvm.cuny.edu}
\author{Qing Sun}
\affiliation{%
Department of Physics and Center for Nuclear Theory\\
Brooklyn College of the City University of New York\\
Brooklyn, New York 11210
}%

\date{April, 2002}

\begin{abstract}
In a recent work we introduced a nonlocal version of the
Nambu--Jona-Lasinio(NJL) model that was designed to generate a
quark self-energy in Euclidean space that was similar to that
obtained in lattice simulations of QCD. In the present work we
carry out related calculations in Minkowski space, so that we can
study the effects of the significant vector and axial-vector
interactions that appear in extended NJL models and which play an
important role in the study of the $\rho$, $\omega$ and $a_1$
mesons. We study the modification of the quark self-energy in the
presence of matter and find that our model reproduces the behavior
of the quark condensate predicted by the model-independent
relation $\langle\bar qq\rangle_{\rho} = \langle\bar
qq\rangle_0(1-\sigma_N\rho_N/f_{\pi}^2m_{\pi}^2+\cdots)$, where
$\sigma_N$ is the pion-nucleon sigma term and $\rho_N$ is the
density of nuclear matter. (Since we do not include a model of
confinement, our study is restricted to the analysis of quark
matter. We provide some discussion of the modification of the
above formula for quark matter.) The inclusion of a quark current
mass leads to a second-order phase transition for the restoration
of chiral symmetry. That restoration is about 80\% at twice
nuclear matter density for the model considered in this work. We
also find that the part of the quark self-energy that is
explicitly dependent upon density has a strong negative
Lorentz-scalar term and a strong positive Lorentz-vector term,
which is analogous to the self-energy found for the nucleon in
nuclear matter when one makes use of the Dirac equation for the
nucleon. In this work we calculate the nucleon self-energy in
nuclear matter using our model of the quark self-energy and obtain
satisfactory results in agreement with the values of the scalar
and vector nucleon potentials in matter found in either
theoretical or phenomenological studies.
\end{abstract}


\maketitle

\section{INTRODUCTION}

In recent years there has been a great deal of interest in
understanding the properties of quark matter at relatively low
temperature and high density [1-5]. Since it is difficult to study
QCD at finite chemical potential using lattice simulations, the
model of choice for such studies has been the Nambu--Jona-Lasinio
model [6-8]. Of particular interest is the prediction of diquark
condensates and color superconductivity at high densities. (Such
studies may be relevant to the properties of neutron stars.)
Calculations made for dense matter using the NJL model are carried
out in Minkowski space. These calculations are limited in that
they do not include a model of confinement and are, therefore,
unable to provide a comprehensive description of the hadronic
phase present at low density and temperature. However, it is
generally believed that a good deal of information may be gained
by studying quark matter, with a proper study of the hadronic
phase deferred until some future time.

As is well known, the NJL model provides a microscopic dynamical
description of chiral symmetry breaking with the generation of
associated quark vacuum condensates and constituent masses. In the
standard versions of the NJL model [6-8], the constituent quark
mass that is generated in the model is a constant. However, it is
known from lattice simulations of QCD that the constituent mass
goes over to the current quark mass when the quark momentum,
$p^2$, is less than about -2\gev{2} [9]. It is our belief that, if
we are to use the NJL model to study dense matter, it is desirable
to make the model as realistic as possible. To that end, we have
introduced a nonlocal version of the NJL model [10] that is able
to reproduce the Euclidean-space behavior of the quark mass seen
in lattice simulations of QCD [9]. To carry out that program we
have introduced a momentum-dependent $q\bar q$ interaction in the
calculation of the quark self-energy and have separated the
regularization of the model from the specification of that
interaction. (This procedure requires the introduction of
additional parameters into the model.)

Recently, we have seen an attempt to obtain the quark self-energy
in Minkowski space by analytic continuation of a Euclidean-space
form based upon gluon exchange enhanced at small momentum
transfers to simulate confinement [11]. The resulting
Minkowski-space values exhibit resonant-like structures and very
large values of the constituent mass. Such results do not appear
to have any natural physical interpretation.

We note that the quark self-energy may be written as \be
\Sigma(\ksq)= A(\ksq)+ B(\ksq)\slr{k} \ee in vacuum. In matter
there is another four-vector, $\eta^\mu$, that describes the
motion of the matter rest frame. It is useful to put $\eta^2=1$
and to note that, if we work in the matter rest frame, we can put
$\eta^\mu=[1,0,0,0]$. In matter, we have \be \Sigma(\ksq,\eta\cdot
\!k)=A(\ksq,\eta\cdot\!k)+B(\ksq,\eta\cdot\!k)\overline{\slr
k}+C(\ksq,\eta\cdot\!k)\slr\eta\,, \ee where we have found it
useful to introduce the four-vector \be
\overline{k^\mu}=k^\mu-(k\cdot\!\eta)\eta^\mu\,. \ee Note that
$\overline{k^0}=0$ in the matter rest frame. In that frame, we may
write \be \Sigma(k^0, \vec k)=A(k^0, \vec k)-B(k^0, \vec
k)\vec\gamma\cdot\!\vec k+\gamma^0C(k^0, \vec k)\,. \ee As we will
see, $\textit{A}$ and $\textit{B}$ satisfy coupled nonlinear
equations, while $\textit{C}$ may be calculated independently.
(Note that $\textit{A}$ and $\textit{C}$ have the same dimension,
while $\textit{B}$ is dimensionless.)

We first discuss our results for the case in which we neglect the
dependence of $\textit{A}$, $\textit{B}$, and $\textit{C}$ on
$k^0$. We anticipate that the dependence on $k^0$ will be
relatively weak and justify that assumption later in this work. As
a further simplification, we will at first neglect $\textit{B}$
and study the behavior of $A(\vec k,\rho)$, where $\rho$ is the
density of quark matter which is taken to contain equal numbers of
up and down quarks. In this case, the maximum value of $A(\vec
k,\rho)$ is found at $\vec k=0$, leading to a simpler presentation
of our results. We then go on to the consideration of the coupled
equations for $A(\vec k,\rho)$ and $B(\vec k,\rho)$. As usual, we
may introduce a density and momentum-dependent mass defined by \be
M(\vec k,\rho)= \frac{A(\vec k,\rho)}{1-B(\vec k,\rho)}\,. \ee We
provide values of $A(\vec k,\rho)$, $M(\vec k,\rho)$ and $C(\vec
k,\rho)$ in the following sections.

We note that the inclusion of $C(\vec k,\rho)$ precludes the
passage to Euclidean space. The analogous problem arises when one
introduces a finite chemical potential. As we will see, $C(\vec
k,\rho)$ is quite large at finite density and can only be
neglected at very small densities. In vacuum, Lorentz invariance
leads to a relation between $C(\ksq)$ and $B(\ksq)$. However, our
formalism does not maintain Lorentz invariance. (For example, our
regulator depends only upon $|\vec k|$.) Therefore, in the
following we will only calculate the contribution to $C(\vec
k,\rho)$ from the matter, which takes the form of two Fermi seas
of up and down positive-energy quarks with Fermi momentum $k_{F}$.

The Lagrangian of our model is \be {\cal L}=&&\bar q(i\slr
\partial-m^0)q +\frac{G_S}{2}\sum_{i=0}^8[
(\bar q\lambda^iq)^2+(\bar qi\gamma_5 \lambda^iq)^2]\nonumber\\
&&-\frac{G_V}{2}\sum_{i=0}^8[
(\bar q\lambda^i\gamma_\mu q)^2+(\bar q\lambda^i\gamma_5 \gamma_\mu q)^2]\nonumber\\
&& +\frac{G_D}{2}\{\det[\bar q(1+\gamma_5)q]+\det[\bar
q(1-\gamma_5)q]\} \nonumber\\
&&+ {\cal L}_{\mbox{\tiny{conf}}}\,. \ee This Lagrangian was used
in Ref. [10] with $G_V=0$. There, both the condensate values and
self-energies were given for the up (down) and strange quarks. For
the present work we neglect the 't Hooft interaction $(G_D=0)$ and
our model of confinement. We also drop the strange quark from
consideration, so that the quark current mass matrix is
$m^0=\mbox{\,diag}(m_u^0, m_d^0)$ with $m_u^0=m_d^0$. Since the 't
Hooft interaction contributes to the self-energy [10], its neglect
leads us to use a larger value of $G_S$ than that we would use if
the 't Hooft interaction were included in our analysis.

The organization of our work is as follows. In Section II we
present the coupled nonlinear equations that determine $A(\vec
k,\rho)$ and $B(\vec k,\rho)$ and also provide an expression for
$C(\vec k,\rho)$, and the density-dependent condensate
$\langle\bar uu\rangle_\rho=\langle\bar dd\rangle_\rho$. We
present values of $A(\vec k,\rho)$ and the condensate, when we
neglect $B(\vec k,\rho)$. In Section III we consider finite values
of $B(\vec k,\rho)$ and provide results for $A(\vec k,\rho)$,
$M(\vec k,\rho)$, $C(\vec k,\rho)$ and the condensate. In Section
IV, we consider the dependence of $\textit{A}$, $\textit{B}$ and
$\textit{C}$ on $k^0$ and provide some results for $A(k_0, \vec
k)$ in vacuum. In Section V we use the quark self-energy
calculated here to obtain the nucleon self-energy in nuclear
matter in a simple model. Finally, Section VI contains some
further discussion and conclusions.

\section{the quark self-energy}

 \begin{figure}
 \includegraphics[bb=0 0 400 400, angle=-2, scale=0.7]{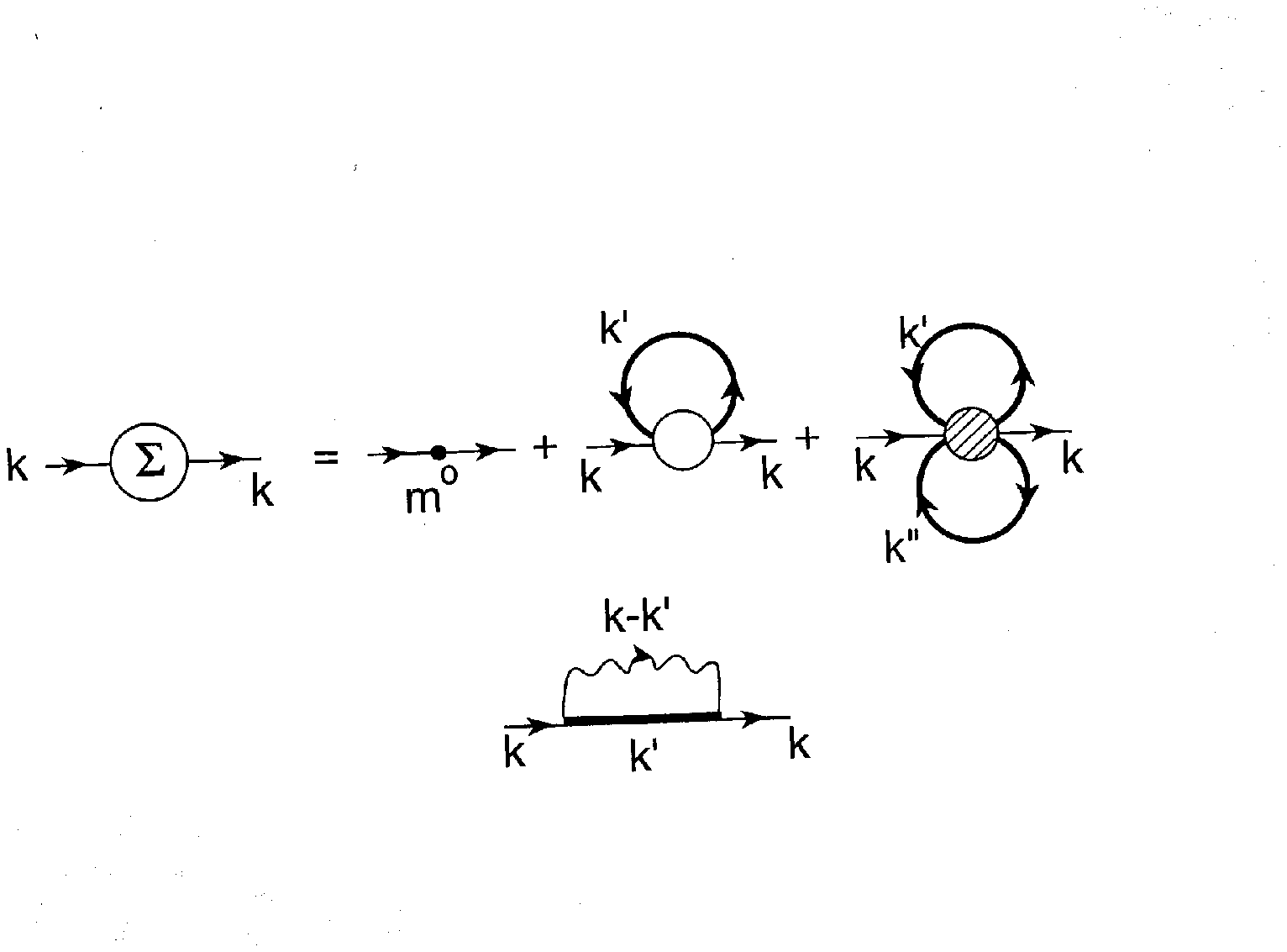}%
 \caption{The equation for the quark self-energy that was solved in Euclidean
 space in Ref. [10] is shown. Here, $m^0$ is the current quark mass. The
 open circle represents the momentum-dependent $q\bar q$ interaction
 of the nonlocal model. The third term on the right-hand side of the figure
 represents the 't Hooft interaction and the fourth term arises from our
 model of confinement. The heavy lines are quark propagators which include the
 self-energy, $\Sigma(k)$, in their definition. [See Eq. (2.1).]}
 \end{figure}

In our earlier work we obtained the quark self-energy from the
solution of the equation depicted in Fig. 1 [10]. There, the open
circle is a momentum-dependent quark interaction, obtained by the
replacement \be G_S\longrightarrow f(k-\kp)\,G_Sf(k-\kp)\,,\ee
where $k$ and $\kp$ are the quark momenta entering (or leaving)
the interaction. We have used \be
f(k-\kp)=\mbox{exp}[-(k-\kp)\,^{2n}/2\beta]\ee with $n=4$ and
$\beta = 20$\gev{8} in Ref. [10]. The corresponding nonlocal
Lagrangian is given in Ref. [10]. On the right-hand side of Fig.1,
the 't Hooft interaction (third term) and the confinement
interaction (fourth term) are neglected for the purposes of this
work. In the second term we have contributions from the
negative-energy states in vacuum as well as the positive-energy
states at finite density, specified by the quark Fermi momentum,
$k_{F}$, of the up and down quark Fermi seas.

In general, the quark propagator is \be iS(k,
k\cdot\!\eta)=\frac{i}{\slr k-\Sigma(\ksq,
k\cdot\!\eta)+i\epsilon}\,,\ee which we will write as \be iS(k^0,
\vec k)=\frac{i}{(k^0-C(k^0, \vec k))\gamma^0-(1-B(k^0,\vec
k))\vec \gamma\cdot\!\vec k-A(k^0, \vec k)+i\epsilon}\,,\ee in the
matter rest frame. In a first approximation we write \be iS(k^0,
\vec k)=\frac{i}{(k^0-C(\vec k))\gamma^0-(1-B(\vec k))\vec
\gamma\cdot\!\vec k-A(\vec k)+i\epsilon}\,,\ee with $A(\vec k)$,
$B(\vec k)$ and $C(\vec k)$ density-dependent, in general. (On
occasion we will write $A(\vec k, \rho)$, etc.) We see that the
presence of $C(\vec k)$ precludes the passage to Euclidean space
that was made in Ref.[10]. That is analogous to the problem
created by the introduction of a chemical potential in the
formalism. Note that \be (k^0-C(\vec k))^2-(1-B(\vec k))^2\vec
k^2-A^2(\vec k)+i\epsilon \nonumber\\
=[k^0-E^+(\vec k)+i\epsilon][k^0-E^-(\vec k)-i\epsilon]\ee with
\be E^+(\vec k)=C(\vec k)+\sqrt{\vec k^2(1-B(\vec k))^2+A^2(\vec
k^2)}\ee and \be E^-(\vec k)=C(\vec k)-\sqrt{\vec k^2(1-B(\vec
k))^2+A^2(\vec k^2)}\,.\ee Here $E^+(\vec k)$ may be interpreted
as the (on-mass-shell) energy of the positive-energy states, while
$E^-(\vec k)$ refers to the negative-energy states.

In order to simplify the notation somewhat, we write \be iS(k^0,
\vec k)=\frac{i}{\slr\Pi(k^0, \vec k)-A( \vec k)+i\epsilon}\,,\ee
where we have defined a four-component quantity \be \Pi^\mu(k^0,
\vec k)=[k^0-C(\vec k),(1-B(\vec k))\vec k]\,.\ee

We also introduce a scalar quantity \be \rho_S(\vec
k)=iN_c(-1)\myint\kp\frac{4A(\vec
\kp)f^2(k-\kp)}{\Pi^2(k^{\,\prime\,0},\vec\kp)-A^2(\vec\kp)+i\epsilon}\,,\ee
where the minus sign is due to the closed Fermion loop in Fig. 1
and the factor of 4 comes from forming the trace associated with
the closed loop. We see that $\rho_S(\vec k)$ does not depend upon
$k^0$, since we are making an on-shell approximation in the
calculation of $f(k-\kp)$. In vacuum we have \be -iA(\vec
k)=-im^0+(2G_Si)\rho_S(\vec k)\,,\ee or \be A(\vec
k)=m^0-2G_S\rho_S(\vec k)\,,\ee where $\rho_S(\vec k)$ is real and
negative. We remark when $f(k-\kp)=1$, $A(\vec k)\rightarrow m$
and $\rho_S(\vec k)\rightarrow\langle\bar uu\rangle$, so that we
regain the usual result [6-8] \be m_u=m_u^0-2G_S\langle\bar
uu\rangle \,,\ee \be m_d=m_d^0-2G_S\langle\bar dd\rangle \,.\ee
(Note that our $G_S$ is one-half of the $G_S$ defined in Ref.
[7].)

The evaluation of $\rho_S(\vec k)$ proceeds by closing to contour
in the complex $k^0$ plane. In the vacuum we find \be \rho_S(\vec
k)=-2N_c\mytint\kp\frac{f^2(k-\kp)A(\vec\kp)} {\sqrt{\vec
k^{\,\prime2}(1-B(\vec\kp))^2+A^2(\vec\kp)}}\,.\ee We define \be
E(\vec k)=\sqrt{\vec k^2(1-B(\vec k))^2+A^2(\vec k)}\,,\ee and \be
(k-\kp)^2=(E(\vec k)-E(\vec \kp))^2-(\vec k-\vec \kp)^2\,,\ee so
that $f(k-\kp)$ depends only upon $|\vec k|$, $|\vec\kp|$ and the
angle between $\vec k$ and $\vec \kp$.

Integrals such as that in Eq. (2.16) require regularization. For
our calculations we insert a factor $\mbox{exp}[-\vec
k^{\,\prime2}\!/\,\alpha^2]$ with $\alpha=0.60$ GeV. (We have used
the same Gaussian regulator in our calculations of meson spectra,
where we have used $\alpha=0.605$ GeV [12-14]. However, those
calculations included a model of confinement, so that we can not
directly take over the parameters $G_S$, $G_V$ and $G_D$ used in
those works.) We remark that an expression for the
density-dependent condensate $\langle \bar uu\rangle_\rho$ may be
obtained by using Eq. (2.16) with $f(k-\kp)=1$.

In the presence of matter it is useful to separate the propagator
into two parts, one of which will give rise to the explicitly
density-dependent terms. In this regard, it is useful to
generalize Eq. (5.8) of Ref. [6]. We define \be \Lambda^{(+)}(\vec
k)=\frac{E(\vec k)\gamma^0-\vec \gamma\cdot\vec
k(1-B(\vec k))+A(\vec k)}{2A(\vec k)}\,,\\
\Lambda^{(-)}(-\vec k)=\frac{-E(\vec k)\gamma^0-\vec
\gamma\cdot\vec k(1-B(\vec k))+A(\vec k)}{2A(\vec k)}\,,\ee where
$E(\vec k)=[\vec k^2(1-B(\vec k))^2+A^2(\vec k)]^{1/2}$. Then \be
S(k)=&&\frac{A(\vec k)}{E(\vec k)}\left[\frac{\Lambda^{(+)}(\vec
k)}{k^0-E^+(\vec k)} - \frac{\Lambda^{(-)}(-\vec k)}{k^0-E^-(\vec
k)}\right]\\ \nonumber &+&2\pi i\frac{A(\vec k)}{E(\vec
k)}\Lambda^{(+)}(\vec k)\,\theta (k_F-|\vec
k|)\,\delta(k^0-E^+(\vec k))\ee In the limit that $A(\vec
k)\rightarrow m^*$ and $B(\vec k)=C(\vec k)=0$, we have, with \be
E^*(\vec k)=\sqrt{\vec k^2+m^{*2}}\,,\ee \be
S(k)&=&\frac{m^*}{E^*(\vec k)}\left[\frac{\Lambda^{(+)}(\vec
k)}{k^0-E^*(\vec k)}
- \frac{\Lambda^{(-)}(-\vec k)}{k^0+E^*(\vec k)}\right]\\
\nonumber &+&2\pi i\frac{m^*}{E^*(\vec k)}\Lambda^{(+)}(\vec
k)\,\theta (k_F-|\vec k|)\,\delta(k^0-E^*(\vec k))\\ &=&
\frac{\slr k+m^*}{k^2-m^{*2}}+\frac{i\pi}{E^*(\vec k)}(\slr
k+m^*)\,\theta(k_F-|\vec k|)\,\delta(k^0-E^*(\vec k))\,,\ee which
agrees with Eq. (5.8) of Ref. [6].

In the presence of matter, we have \be A(\vec k,
\rho)=m^0-2G_S[\rho_S^{\mbox{\tiny{vac}}}(\vec
k)-\rho_S^{\mbox{\tiny{mat}}}(\vec k)]\,,\ee where
$\rho_S^{\mbox{\tiny{mat}}}(\vec k)$ is calculated in the same
manner as $\rho_S^{\mbox{\tiny{vac}}}(\vec k)$, except that the
upper limit of the integral over $|\vec \kp|$ is $k_{F}$. Equation
(2.21) is a generalization of a corresponding equation that may be
found in Klevansky's review. (See Eqs. (5.18) of Ref. [6].)

If we neglect $B(\vec k)$, Eqs. (2.11) and (2.16) provide a
nonlinear equation for $A(\vec k)$ which may be solved by
iteration. The results of such a calculation are reported in Table
I where, for nuclear matter, we put $k_{{F}}=0.268$ GeV. In Figs.
2 and 3 we show values of the condensate $\langle \bar uu\rangle$
and $A(0,\rho)$ as a function of the density. These results may be
usefully discussed in terms of the relation [15]\be\langle \bar
uu\rangle_\rho=\langle \bar
uu\rangle_0\left(1-\frac{\sigma_N\rho_N}{f_\pi^2m_\pi^2}+\cdots\right)\,.\ee
Here $\sigma_N$ is the pion-nucleon sigma term and $\rho_N$ is the
density of nucleons. If we put $\sigma_N=0.050$ GeV,
$\rho_N=(0.109\,\mbox{GeV})^3$, $f_\pi=0.0942$ GeV and
$m_\pi=0.138$ GeV, we find a 38\% reduction of the condensate at
nuclear matter density, which agrees with our results given in
Table I and Fig. 2. It is of interest to note that the linear
dependence on the density implied by Eq. (2.26) appears to be
valid up to about twice nuclear matter density. However, one may
be concerned that, since we study quark matter rather than nuclear
matter, Eq. (2.26) may not be appropriate. Consider, however, the
relation \be\langle \bar uu\rangle_\rho=\langle \bar
uu\rangle_0\left(1-\frac{\sigma_q\rho_q}{f_\pi^2m_\pi^2}+\cdots\right)\,,\ee
where $\sigma_q$ is the quark sigma term and $\rho_q$ is the
number density of the quarks, which we may put equal to $3\rho_N$.
Thus, if $3\sigma_q=\sigma_N$, we may use Eq. (2.26). The fact
that $\sigma_q\simeq15$ MeV has been discussed by Vogl and Weise
[7]. We have also discussed this matter in great detail in Ref.
[16], where we calculated similar values of $\sigma_q$ using the
standard version of the NJL model. We conclude that the use of Eq.
(2.26), or Eq. (2.27), with an appropriate value of $\sigma_q$, is
satisfactory. For example, if $\sigma_q=\sigma_N/3$, as suggested
in Ref. [7], the two relations imply the same density dependence
of the condensate.

In Fig. 3 we show $A(0, \rho)$ which is the density-dependent mass
parameter of the theory when $B(\vec k,\rho)=0$. We see that $A(0,
\rho)$ follows the trend seen in Fig.2 for the density dependence
of the condensate.

 \begin{figure}
 \includegraphics[bb=0 0 400 200, angle=-0.5, scale=1]{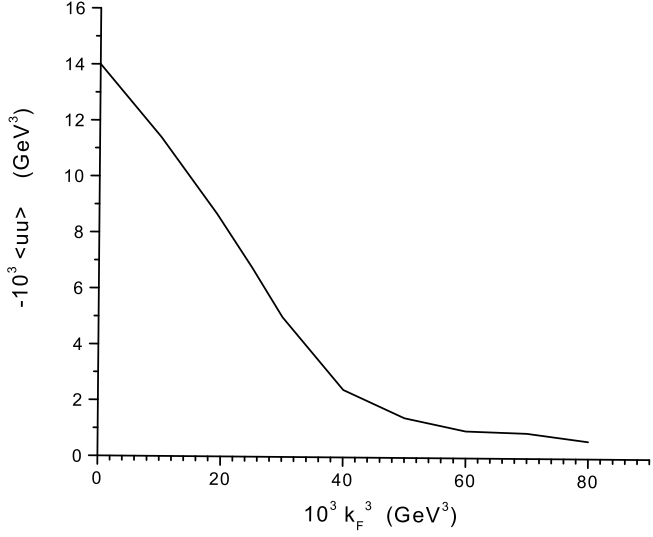}%
 \caption{Values of the condensate $\langle\bar uu\rangle$ are given as a
 function of $10^3k_{F}^3$. For nuclear matter $10^3k_{F}^3=19.2$\gev 3. [See Table I.]
 Here $G_S=13.0$\gev{-2} and $B(\vec k,\rho)$ is put equal to zero.}
 \end{figure}

 \begin{figure}
 \includegraphics[bb=0 0 400 250, angle=-0.5, scale=1]{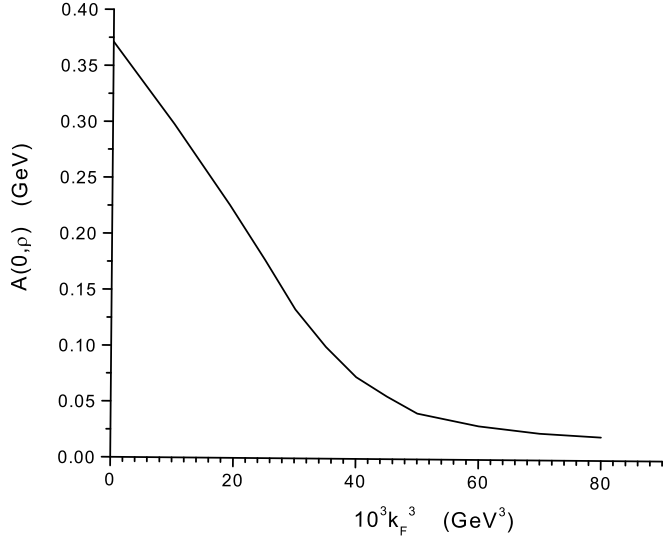}%
 \caption{Values of $A(0,\rho)$ are given as a function of $10^3k_{F}^3$.
 [See Table I and the caption of Fig. 2.]}
 \end{figure}

\begin{table}
 \begin{tabular}{||@{\hspace{0.5cm}}
 c@{\hspace{0.5cm}}|@{\hspace{0.5cm}}c@{\hspace{0.5cm}}
 |@{\hspace{0.5cm}}c@{\hspace{0.5cm}}|@{\hspace{0.5cm}}c@{\hspace{0.5cm}}||}\hline\hline
 $10^3k_{F}^3$ &$-\langle\bar uu\rangle^{1/3}$ &$-10^3\langle\bar uu\rangle$ &$A(0,\rho)$ \\
 (\gev{3})   &(GeV)                          &(\gev{3})                    &(GeV)\\\hline\hline
 0           &0.241                          &14.0                         &0.371\\\hline
 10          &0.225                          &11.4                         &0.298\\\hline
 19.2(nm)    &0.205                          &8.67                         &0.226\\\hline
 25          &0.189                          &6.75                         &0.177\\\hline
 30          &0.171                          &5.00                         &0.133\\\hline
 40          &0.134                          &2.46                         &0.077\\\hline
 50          &0.113                          &1.44                         &0.041\\\hline
 60          &0.0990                         &0.967                        &0.030\\\hline
 70          &0.0915                         &0.898                        &0.024\\\hline
 80          &0.0853                         &0.620                        &0.021\\\hline
 90          &0.0805                         &0.522                        &0.018\\\hline\hline
 \end{tabular}
 \vspace{1.2cm}
 \caption{Values of the condensate and $A(0,\rho)$ are
  given for $G_S=G_V=13.0$\gev{-2} and $m_u^0=m_d^0=0.005$ GeV. Here
  $k_{F}=0.268$ GeV for nuclear matter. We note the reduction of
  the condensate of 38\% and a 40\% reduction of $A(0,\rho)$ at nuclear matter
  density [$10^3k_{F}^3=19.2$\gev 3]. Here $\alpha=0.60$ GeV is used in the
  Gaussian regulator $\mbox{exp}[-\vec k^2/\alpha^2]$.}
 \end{table}

\section{lorentz-vector terms of the quark self-energy}

We now write $\Sigma(\vec k)=\Sigma_S(\vec k)+\Sigma_V(\vec k)$
where \be \Sigma_V(\vec k)=-\vec\gamma\cdot\vec kB(\vec
k)+\gamma^0C(\vec k)\,.\ee For the calculation of $C(\vec k)$ we
obtain the contribution from the last term in Eq. (2.21), with the
result that \be C(\vec k)=2G_V\rho_2^V(\vec k)\ee with \be
\rho_2^V(\vec
k)=2N_c\!\int^{k_{F}}\!\!\!\frac{d^3\!{\kp}}{(2\pi)^3}\,f^{\,2}(k-\kp)\,.\ee
An expression for $B(\vec k)$ may be found from the relation \be
-i[-\vec\gamma\cdot\vec kB(\vec k)]=(-2G_Vi)(-1)i\myint \kp
S(\kp)f^{\,2}(k-\kp)\ee if we only keep the term proportional to
$\vec \gamma\cdot\vec\kp$ in the expression for the quark
propagator. In vacuum we may compare corresponding terms in Eq.
(3.4) \be -\vec\gamma\cdot\vec kB(\vec
k)=-2G_V\myint\kp\frac{-\vec\gamma\cdot\vec \kp[1-B(\vec
\kp)]f^{\,2}(k-\kp)}{[\kp_0-E^+(\vec\kp)+i\epsilon][\kp_0-E^-(\vec\kp)-i\epsilon]}\,.\ee
Thus, \be B^{\mbox{\tiny{vac}}}(\vec
k)=2G_V\rho_1^{\mbox{\tiny{vac}}}(\vec k)\,,\ee with \be |\vec
k|\rho_1^{\mbox{\tiny{vac}}}(\vec k)=2N_c\mytint\kp\frac{|\vec
\kp|(\hat k\cdot\hat\kp)f^{\,2}(k-\kp)[1-B(\vec\kp)]}{\sqrt{\vec
k^{\,\prime2}[1-B(\vec\kp)]^2+A^2(\vec \kp)}}\,.\ee Here, $\hat k$
and $\hat\kp$ are unit vectors. As in the calculation of $A(\vec
k)$, Eq. (3.6) is generalized to read \be B(\vec
k)=2G_V[\rho_1^{\mbox{\tiny{vac}}}(\vec
k)-\rho_1^{\mbox{\tiny{mat}}}(\vec k)]\,,\ee where
$\rho_1^{\mbox{\tiny{mat}}}(\vec k)$ is calculated using Eq. (3.7)
with an upper limit on $|\vec\kp|$ of $k_{F}$.

In Table II we present results of our calculation of the
condensate, $A(0,\rho)$, $A(0,\rho)/[1-B(0,\rho)]$, $C(0,\rho)$,
$B(0,\rho)$ and $A(0, \rho)-A(0,0)$. We also define \be U_S(\vec
k, \rho)=A(\vec k, \rho)-A(\vec k, 0)\,,\ee where $ U_S(\vec k,
\rho)$ is the density-dependent modification of $A(\vec k, 0)$ in
matter.

It may be seen from the values given in Table II that there is a
thirty percent reduction of the condensate at the density of
nuclear matter, while the value of $A(0)$ is reduced by
thirty-nine percent. We would obtain a thirty percent reduction of
the condensate if $\sigma_N=39$ MeV.

In Figs. 4 and 5 we exhibit values of $A(\vec k,\rho)$ and $A(\vec
k,\rho)/[1-B(\vec k, \rho)]$ for various densities and in Fig. 6
we present values of $C(\vec k,\rho)$. Figure 7 shows the values
of $U_S(\vec k,\rho_{NM})$ and $C(\vec k,\rho_{NM})$.

\begin{table}
 \begin{tabular}{||@{\hspace{0.2cm}}c@{\hspace{0.2cm}}|@{\hspace{0.2cm}}c@{\hspace{0.2cm}}
 |@{\hspace{0.2cm}}c@{\hspace{0.2cm}}|@{\hspace{0.2cm}}c@{\hspace{0.2cm}}|@{\hspace{0.2cm}}
 c@{\hspace{0.2cm}}|@{\hspace{0.2cm}}c@{\hspace{0.2cm}}|@{\hspace{0.2cm}}c@{\hspace{0.2cm}}
 |@{\hspace{0.2cm}}c@{\hspace{0.2cm}}||}\hline\hline
 $10^3k_{F}^3$ &$-\langle\bar uu\rangle^{1/3}$ &$-10^3\langle\bar uu\rangle$ &$A(0,\rho)$
 &$\frac{A(0,\rho)}{1-B(0,\rho)}$ &$U_S(0,\rho)$ &$C(0,\rho)$ &$B(0,\rho)$\\
 (\gev{3})   &(GeV)     &(\gev{3})    &(GeV)    &(GeV)    &(GeV)    &(GeV)    &       \\\hline\hline
 0           &0.2401    &13.85        &0.365    &0.347    &0        &0        &-0.0544\\\hline
 10          &0.2279    &11.84        &0.292    &0.272    &-0.073   &0.0794   &-0.0736\\\hline
 19.2        &0.2128    &9.64         &0.223    &0.203    &-0.142   &0.0988   &-0.1019\\\hline
 30          &0.1755    &5.36         &0.114    &0.0942   &-0.252   &0.115    &-0.210 \\\hline
 40          &0.1441    &2.99         &0.0606   &0.0435   &-0.304   &0.126    &-0.394 \\\hline
 50          &0.1291    &2.15         &0.0432   &0.0279   &-0.322   &0.136    &-0.545 \\\hline\hline
 \end{tabular}
 \vspace{1.2cm}
 \caption{Various values are given for the case $G_S=13.5$\gev{-2}, $G_V=10.0$\gev{-2},
  $m_u^0=m_d^0=0.005$ GeV and $\alpha=0.60$ GeV. Note a reduction of 30\% for the condensate
  and 39\% for $A(0,\rho)$ at nuclear matter density, where $k_{F}=0.268$ GeV and $10^3k_{F}^3
  =19.2$\gev 3.}
 \end{table}

 \begin{figure}
 \includegraphics[bb=0 0 400 200, angle=-0.5, scale=1]{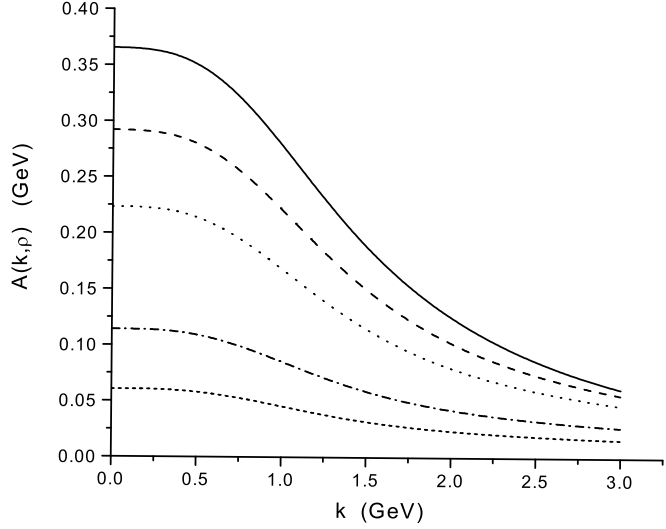}%
 \caption{Values of $A(\vec k, \rho)$ are given as a function of $|\vec k|$
 for various densities: a) $10^3k_{{F}}^3=0$ [solid line]; b) $10^3k_{{F}}^3=10.0$\gev 3
 [dashed line]; c) $10^3k_{{F}}^3=19.2$\gev 3 [dotted line]; d) $10^3k_{{F}}^3=30.0$\gev 3
 [dot-dash line] and e) $10^3k_{{F}}^3=40.0$\gev 3 [short dash]. Here $G_S=13.5$\gev{-2},
 $G_V=10.0$\gev{-2}, $m^0=0.005$ GeV and $\alpha=0.60$ GeV. [See Table II.]}
 \end{figure}

 \begin{figure}
 \includegraphics[bb=0 0 400 250, angle=-0.5, scale=1]{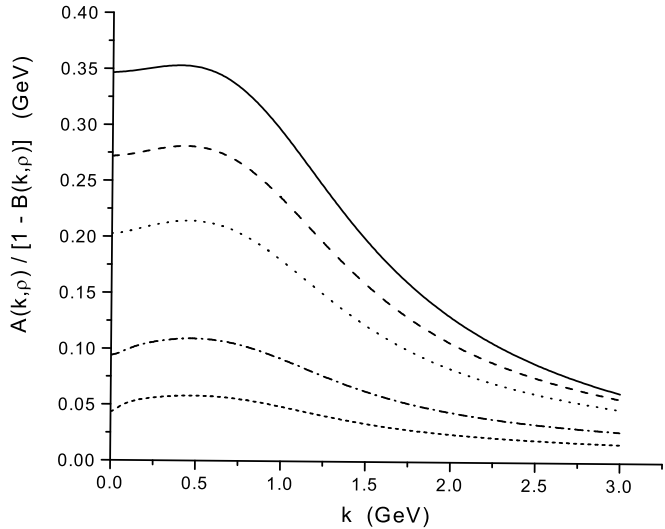}%
 \caption{The quantity $A(\vec k,\rho)/[1-B(\vec k,\rho)]$, which plays
 the role of a momentum- and density-dependent mass parameter, is shown. [See Table
 II and the caption of Fig. 4.]}
 \end{figure}

 \begin{figure}
 \includegraphics[bb=0 0 400 250, angle=-0.5, scale=1]{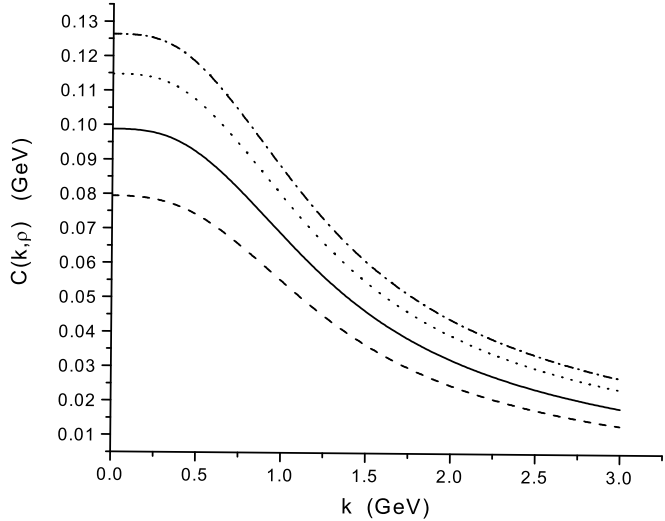}%
 \caption{Values of $C(\vec k, \rho)$ are shown for various densities:
 a) $10^3k_{F}^3=10.0$\gev 3 [dashed line]; b) $10^3k_{F}^3=19.2$\gev 3 [solid line];
 c) $10^3k_{F}^3=30.0$\gev 3 [dotted line] and d) $10^3k_{F}^3=40.0$\gev 3
 [dash-dot line]. Here $G_S=13.5$\gev{-2}, $G_V=10.0$\gev{-2},
 $m^0=0.005$ GeV and $\alpha=0.60$ GeV. }
 \end{figure}

 \begin{figure}
 \includegraphics[bb=0 0 400 250, angle=-0.5, scale=1]{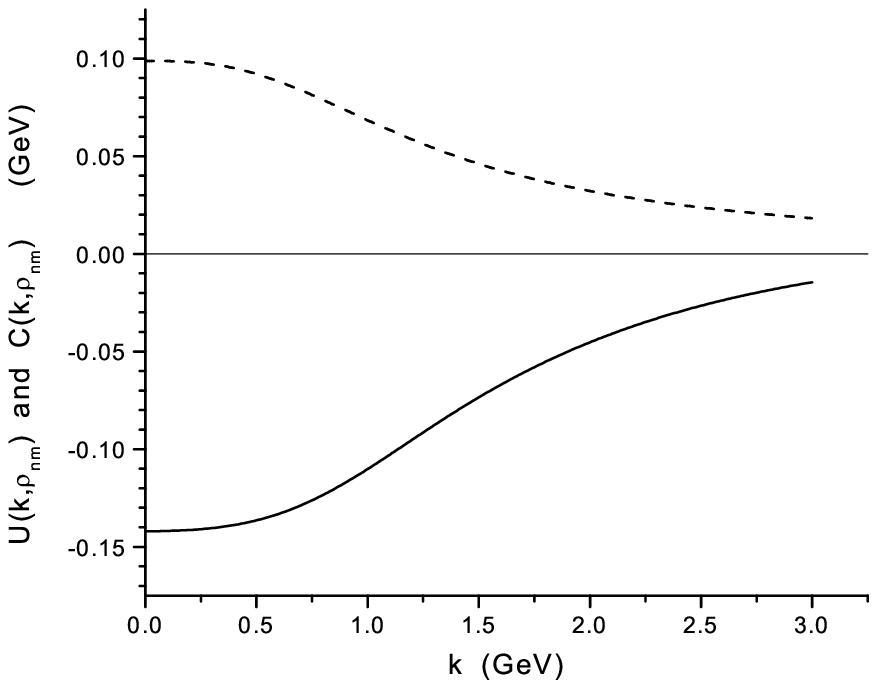}%
 \caption{The values of $U(\vec k, \rho_{NM})=A(\vec k, \rho_{NM})-A(\vec k, 0)$
 [solid line] and $C(\vec k, \rho_{NM})$ [dashed line] are shown. $U(\vec k, \rho_{NM})$
 represents the density-dependent correction to the vacuum value of the
 scalar term of the quark self-energy.}
 \end{figure}

\section{off-mass-shell effects}

In our calculations we have neglected the $k_0$ dependence of
$\textit{A}$, $\textit{B}$ and $\textit{C}$. In this Section we
provide some justification of that approximation. We can combine
Eqs. (2.9) and (2.11) to yield a nonlinear equation for
$A(k^0,\vec k)$ in vacuum \be A(k^0,\vec
k)=-(2G_Si)N_c(-1)\myint\kp
\frac{4A(k^{\,\prime0},\vec\kp)f^{\,2}(k-\kp)}{\slr k^{\,\prime}
-A(k^{\,\prime0},\vec\kp)+i\epsilon}\,.\ee Here we have neglected
$m^0$ and $B(k^0,\vec k)$. We see that $k^0$ dependence arises
from $f^{\,2}(k-\kp)$ which is given by \be
f^{\,2}(k-\kp)=\mbox{exp}[-(k-\kp)^{\,2n}/\beta]\,,\ee with \be
(k-\kp)^2=(k^0-k^{\,\prime0})^2-(\vec k-\vec \kp)^2\,,\ee which
differs from the on-mass-shell version given in Eq. (2.18).

The solution of Eq. (4.1) for $A(k^0, \vec k)$ is shown in Fig. 8
for various values of $k^0$. It is seen that the dependence on
$k^0$ is weak as was assumed in this work.

 \begin{figure}
 \includegraphics[bb=0 0 400 200, angle=-0.5, scale=1.5]{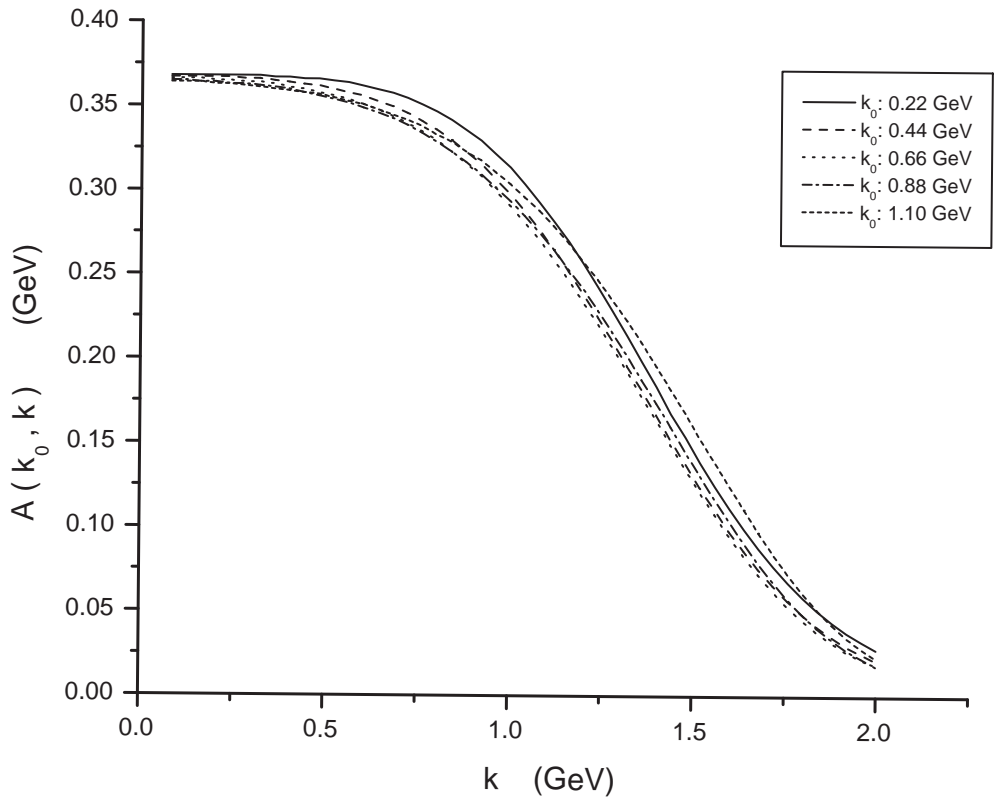}%
 \caption{Values of $A(k^0, \vec k)$ are shown as a function of $|\vec k|$ for
 various values of $k^0$.}
 \end{figure}

\section{the nucleon self-energy in matter}

If one uses the Dirac equation to describe the interaction of a
nucleon with a nucleus, or with nuclear matter, it is found that a
strong scalar attraction is needed as well as a strong vector
repulsion [17, 18]. The scalar field is of the order of -400 MeV
and the vector field is about 300 MeV. It is of interest to see if
the nucleon self-energy, $\Sigma_N=V_S+\gamma^0V_V$, can be
calculated in terms of the quark self-energy obtained in this
work. To carry out this program we use a simple model of the
nucleon in which a quark is coupled to a scalar diquark. The full
complexity of the wave function, including vector diquarks and
various relativistic effects, is discussed in Ref. [19].

We can calculate the nucleon self-energy in nuclear matter using a
triangle diagram in which one of the lower two vertices of the
triangle represents a vertex function for a zero-momentum nucleon
to emit a quark of momentum $\vec k$ leaving a spectator
(on-mass-shell) diquark of momentum $-\vec k$. The other lower
vertex represents the inverse process. At the upper vertex we
insert the quark self-energy, $\Sigma(\vec k)=U_S(\vec
k)+\gamma^0C(\vec k)$ calculated in this work and integrate over
$\vec k$. We make use of Fig. 4 of Ref. [19] and parametrize the
product of the vertex function and the quark Greens function by
the quark-diquark wave function \be \psi(\vec k)=\frac{1}{\sqrt
N}\, e^{-\vec k^2/\lambda^2}u(\vec k,s)\,,\ee where $u(\vec k, s)$
is a spinor for a quark of momentum $\vec k$ and spin projection
$s$ [19].

The normalization for a single quark is obtained from the relation
\be \frac{1}{N}\mytint ke^{-2\vec
k^2/\lambda^2}\left(1+\frac{{\vec k}^2}{[E_q(\vec
k)+m_q]^2}\right)=1\,,\ee where we put $\lambda=0.18$ GeV to
correspond to the results of Ref. [19]. We may then relate the
nucleon self-energy to the quark self-energy. For a nucleon of
momentum $\vec P=0$, we have, with $E_q(\vec k)=[\vec
k^2+m_q^2]^{1/2}$, and $m_q=0.364$ GeV, \be V_V=\frac{3}N\mytint k
C(\vec k)\,e^{-2\vec k^2/\lambda^2}\left(1+\frac{{\vec
k}^2}{[E_q(\vec k)+m_q]^2}\right)\,,\ee and \be V_S=\frac3N\mytint
k U_S(\vec k)\,e^{-2\vec k^2/\lambda^2}\left(1-\frac{{\vec
k}^2}{[E_q(\vec k)+m_q]^2}\right)\,.\ee The difference of sign in
the brackets appearing in Eqs. (5.3) and (5.4) is due to the
different behavior of the Dirac matrices, $\mathbf {1}$ and
$\gamma^0$, at the upper vertex of the triangle.

Since the momentum content of the quark-diquark wave function is
small [19], we expect that $V_S\simeq 3U_S(0)$ and $V_V\simeq
3C(0)$, so that $V_V\simeq 296$ MeV and $V_S\simeq -426$ MeV. A
more careful evaluation of the integrals in Eqs. (5.3) and (5.4)
yields $V_V=295$ MeV and $V_S=-392$ MeV, which is in general
accord with the values given in Refs. [17] and [18].

\section{discussion}

The behavior of matter at high density and low temperature has
received a good deal of attention in the last few years [1-5]. A
large part of the work in this area has made use of the NJL model.
In our work we have attempted to modify the NJL model so that its
predictions are in greater accord with QCD. As a first step, in
Ref. [10] we introduced a momentum-dependent $q\bar q$ interaction
which allowed us to reproduce the Euclidean-space behavior for the
constituent quark mass obtained in lattice simulation of QCD [9].
In the present study we have considered the important vector
interactions of an extended NJL model. Of particular interest is
the behavior of the quark condensate at finite density. We find
that the linear behavior in the density exhibited in Eq. (2.26)
holds in our model up to about twice nuclear matter density. Also,
we note that the use of a nonzero current quark mass is important
for that result, since in the absence of an explicit chiral
symmetry breaking term, the model exhibits a first-order phase
transition at about 1.25 times nuclear matter density.

Another point of interest are the results shown in Fig. 7. The
quark self-energy is similar to that found in relativistic nuclear
physics, with strong scalar attraction and strong vector
repulsion. The simple calculation reported in Section V suggests
that the quark self-energy, when multiplied by 3, provides a
satisfactory estimate of the nucleon self-energy which is in
accord with results of relativistic nuclear physics [17, 18].

There is a body of work based upon the solutions of the
Schwinger-Dyson and Bethe-Salpeter equations with a
phenomenological form for the gluon propagator [20-23]. This body
of work is reviewed in Ref. [24]. In contrast to the results of
our work, it is found that the quark condensate
$\textit{incre{as}es}$ with increasing chemical potential. The
authors of Ref. [20] argue that the baryon density is zero up to
the critical chemical potential, $\mu_c$, for deconfinement. They
state that ``This result is an expected consequence of confinement
which entails that each additional quark must be locally paired
with an antiquark thereby increasing the density of condensate
pairs as $\mu$ is increased. For this reason, as long as
$\mu<\mu_c$, there is no excess of particles over antiparticles in
the vacuum and hence the baryon number density remains zero." We
note, however, that the baryon density usually considered in such
calculations is due to the presence of $\textit{nucleons}$, which
have a finite baryon density. As noted earlier, in our work the
baryon density is due to the presence of Fermi seas of up and down
quarks, which serve to provide a model of the baryon density that
would arise due to the presence of nucleons. We suggest that the
conclusions presented in Ref. [20] do not refer to the situation
of physical interest in which one studies the value of the quark
condensate in nuclei or in nuclear matter.


\newpage
\vspace{1.5cm}
\noindent$\textbf{References}$\\[-2cm]


\end{document}